\documentclass{article}
\usepackage{amsmath, amsfonts, amssymb, amsthm, bbm}
\usepackage{graphicx}

\begin{document}
\title{Effective Hamiltonian of Liquid-Vapor Curved Interfaces
in Mean Field}

%
%
%

\author{Jos\'e~G.~Segovia-L\'opez$^{1}$\thanks{jose.segovia@dacb.ujat.mx},
$$ Adolfo~Zamora$^{2}$\thanks{zamora@correo.cua.uam.mx}
$$ and Jos\'e Antonio Santiago$^{2}$\thanks{jsantiago@correo.cua.uam.mx}
\\[0.5cm]
$^{1}$\it Divisi\'on Acad\'emica de Ciencias B\'asicas,\\
\it Universidad Ju\'arez Aut\'onoma de Tabasco,\\
\it Km 1 Carretera Cunduac\'an-Jalpa, Apartado Postal 24,\\
\it 86690, Cunduac\'an, Tabasco, Mexico
\\[0.5cm]
$^{2}$\it Departamento de Matem\'aticas Aplicadas y Sistemas,\\
\it Universidad Aut\'onoma Metropolitana -- Cuajimalpa,\\
\it M\'exico D.F. 01120, Mexico
\\[0.3cm]} 

\date{}
\maketitle

\begin{abstract}
We analyze a one-component simple fluid in a liquid-vapor
coexistence state, which forms an arbitrarily curved interface. By
using an approach based on density functional theory, we obtain an
exact and simple expression for the grand potential at the level
of mean field approximation that depends on the density profile
and the short-range interaction potential. By introducing the
step-function approximation for the density profile, and using
general geometric arguments, we expand the grand potential in
powers of the principal curvatures of the surface and find
consistency with the Helfrich phenomenological model in the second
order approximation.
\end{abstract}



\maketitle

\newpage

\section{Introduction}
\label{sn:Intro}

\noindent
Although the description of a one-component fluid in a liquid-vapor
coexistence state has been studied since long ago, it still remains
as a topic of interest for inhomogeneous fluids~%
\cite{evans, widom, tolman, dietrich}. Not just open questions are
yet to be answered within this problem, but also its appropriate
description may be useful for the analysis of more complex systems
such as colloids, polymers, mixtures of these, etc. Particularly,
investigation of the microscopic expression for the free energy
that represents an arbitrarily curved interface remains a topic of
interest for the one-component fluid in a coexistence state. In
order to analyze this system different approximation schemes,
rigorous derivations, and many genuine shortcuts have been
implemented~\cite{dietrich,Triezenberg,gibbs,blokhouis,Percus}.
Associated to the free energy there is the calculation of
microscopic expressions for interfacial properties such as surface
tension, spontaneous curvature, and rigidity coefficients. The
main objective of this work is to derive the free energy and all
these quantities in a rigorous and fully general form.\\

One of the approximation schemes that has provided a first principles
description is the stress tensor theory, implemented initially by
Romero-Roch\'{\i}n, Varea, and Robledo~\cite{romero}. In this theory
the authors identify the normal component of the stress tensor as the
fundamental quantity to calculate the grand potential, which represents
the energetic cost to maintain the interface having a given geometry.
Construction of the most general form of this stress tensor is due to
Percus and Romero-Roch\'{\i}n, and represents one of the important
achievements within this approximation scheme~\cite{percus, romeropercus}.
The microscopic expression for this stress tensor has been used to
describe the interfacial region, within the van der Waals model, when
the interface has planar, spherical, and cylindrical geometries. For
each of these interfaces, the exact microscopic expression for the grand
potential has been constructed considering drops of arbitrary size~%
\cite{segovia}. Nevertheless, in order to compare these results, it is
convenient to approximate the grand potential as a limiting case; which
is, when the radius of the Gibbs dividing surface is much larger than
the range of the interaction potential. This has lead to a microscopic
expression for the grand potential as a power series of the inverse
radii of curvature. At this level, we establish an equivalence between
the microscopic expression for the grand potential and the Helfrich
phenomenological model~\cite{Helfrich} for each of the corresponding
geometries.\\

Despite the Helfrich model is widely used to describe a variety of
systems~\cite{blokhouis, robledo1}, no first-principles derivation
of the model is known. Several efforts have been made aimed at
achieving the task~\cite{robledo}. The usual strategy to get to the
phenomenological form of the Helfrich free energy is based on the
assumption that the interface is a bidimensional elastic
continuous-medium~\cite{Landau-elas}. According to this model, the
free energy of the interface is a function of the principal
curvatures of the system, which can be written as
\begin{equation}
\Omega_{S}=\int dS[\gamma-2\kappa c_{0}H+\kappa H^{2}+ \bar\kappa K],
\label{hel}
\end{equation}
where $H=(1/R_{1})+(1/R_{2})$ and $K=1/R_{1}R_{2}$ are the mean and
Gaussian curvature respectively, with$R_{1}$ and $R_{2}$ being the
principal radii of curvature. The coefficients $\gamma $, $c_{0}$,
$\kappa$, and $\bar\kappa$ are equilibrium surface properties.\\

As a second part of this work, we carry out a
rigorous derivation of the Helfrich model corresponding to a fluid
interface, Eq. (\ref{hel}), calculate the surface properties, and
show that it is a legitimate model to describe arbitrarily curved
interfaces. We follow an essentially different program as compared
to previous approaches. Instead of analyzing diverse particular
geometries, which is not a simple task, we consider the general
case of an arbitrarily curved interfacial region and study it
using a semi-orthogonal coordinate system. A notable fact that
appears in the description of curved interfaces concerns the
localization of the Gibbs dividing surface. As the physical
properties of the system must be independent of this choice, a
displacement in the localization of the interface in microscopic
distances shall not modify the value of the surface tension, but
will alter the value of the rigidity coefficients. Thus, different
localizations of the Gibbs dividing surface give rise to different
values of the rigidity coefficients, which results in an arbitrariness
of their values. Explanation of the origin of this arbitrariness is
one of the questions yet to be addressed. In this study, we fix
the radius of the Gibbs dividing surface within the approximation
introduced for the density profile. Although this criterion is not
unique, the results obtained are consistent with previous works.\\

This paper is organized as follows. In Section \ref{sn:S-T-T} we
briefly outline the stress tensor formalism that describes a smooth,
but otherwise general, interface. Section \ref{sn:F-E-C-I} is devoted
to the calculation of the microscopic expression for the grand
potential of a curved interface within the van der Waals approximation.
Next, the limit of large radii of curvature, as compared to the range
of the interaction potential, is introduced in Section \ref{sn:L-A-I}
and a further approximation to the grand potential is obtained, which
coincides with the Helfrich phenomenological model. Finally, some
concluding remarks are drawn in Section \ref{sn:concl}.\\

\section{Stress Tensor Theory}
\label{sn:S-T-T}

According to density functional theory, the grand potential can be
written in the form~\cite{evans, widom}
\begin{equation}
\Omega[\rho(\vec{r})]= F[\rho(\vec{r})]+ \int d\vec{r}[\mu
-V_{\rm ext}(\vec{r})]\rho(\vec{r}),
\end{equation}
where $F[\rho(\vec{r}\,)]$ is the intrinsic Helmholtz free energy,
$\mu$ is the chemical potential, and $V_{\rm ext}$ is the external potential.
The equilibrium value of the density profile is obtained by minimizing
the grand potential; that is by solving equation
\begin{equation}
\left.{\delta F[\rho(\vec{r})]\over \delta \rho}\right|_{\rho_0}
-[\mu - V_{\rm ext}(\vec{r})]=0.
\label{ecfund}
\end{equation}
In absence of exact analytic solutions, the usual approach is through
numerical methods, which provide estimate values. However, in this work
we insist in an analytic solution and follow the route of the stress
tensor to obtain it. Next we briefly outline the general formalism,
which can be read in detail in Ref. \cite{romero}.\\

In an inhomogeneous fluid the condition for mechanical equilibrium
implies the existence of a conservation equation, obtained from
(\ref{ecfund}), which may be expressed as
\begin{equation}
\nabla \cdot \sigma =\rho_0 \nabla V_{\rm ext},
\label{consec}
\end{equation}
with $\sigma$ being the stress tensor and $\rho_0 V_{\rm ext}(\vec{r})$
the external force per unit area.\\

The stress tensor is not unique. As it can be seen from (\ref{consec}),
a term with zero divergence may always be added. Nature of the system
suggests a separation of the stress tensor in two contributions
\begin{equation}
\sigma=\sigma_{0}+\sigma_{S},
\end{equation}
where $\sigma_{0}$ is the homogeneous part and $\sigma_{S}$ is the
inhomogeneous one. The homogeneous part describes the bulk phases, where
the density profile is uniform, whereas the inhomogeneous region is that
where $\nabla \rho_0 (\vec{r}) \neq 0$.\\

The free energy of the system is obtained by integrating the normal
component of the stress tensor over the whole space. Once again, we
separate this normal component in two contributions, one associated to
the homogeneous region and the other from the inhomogeneous one~%
\cite{romero}
\begin{equation}
\Omega[\rho_0(\vec{r})]= \int d\vec{r} \sigma^{N}(\vec{r})
= \int d\vec{r}\sigma_{0} ^{N}(\vec{r})-\int d\vec{r}
\sigma^{N}_{S}(\vec{r}).
\end{equation}
As we are interested in the inhomogeneous region, we neglect the
contribution to the energy arising from the bulk phases. From now on
we concentrate in obtaining the microscopic expression for $\Omega_{S}$,
which is given by
\begin{equation}
\Omega_{S}[\rho_0(\vec{r})]= -\int d\vec{r}\sigma^{N}_{S}(\vec{r}).
\label{sup}
\end{equation}
We observe that $\sigma_{S}^{N}$ is the fundamental quantity
needed to obtain microscopic expressions for the properties of the
surface. For the system under study the microscopic expression for
the stress tensor of the inhomogeneous region, within the van der
Waals approximation, is~\cite{romeropercus, segovia}
\begin{eqnarray}
\sigma^{\alpha\beta}_{S}(\vec{r}) &=& -\int d\vec{r}\,'\int^{1}_{0}
d\lambda\rho_0(\vec{r}-(1-\lambda)\vec{r}\,') \tilde{\omega}(|\vec{r}\,'|)
r'_{\alpha}\nabla_{\beta} \rho_0(\vec{r}+\lambda\vec{r}\,')\nonumber
\\ &-& \frac{1}{2}\nabla_{\nu}\int d\vec{r}\,'\int^{1}_{0}
d\lambda\rho_0(\vec{r}-(1-\lambda)\vec{r}\,')\tilde{\omega} (|\vec{r}\,'|)
\nonumber \\ &\times& r'_{\beta}
\left[r'_{\alpha}\nabla_{\nu}\rho_0(\vec{r}+\lambda\vec{r}\,')
-r'_{\nu}\nabla_{\alpha}\rho_0(\vec{r}+\lambda\vec{r}\,')\right].
\label{cuatro}
\end{eqnarray}
This quantity depends exclusively on the density profile, the
interaction potential, and is independent of the geometry of the
interface. In fact, the geometry is defined by the density
profile. For instance, for a planar interface $\rho_0(\vec{r})=
\rho_0(z)$, and for a spherical interface $\rho_0(\vec{r})=
\rho_0(|\vec{r}|)$. In general, for an arbitrary interface if $\xi$
denotes the normal coordinate, the density profile is a function
exclusively of this quantity: $\rho_0(\vec{r}) = \rho_0(\xi)$. By
using this information, we go on to calculating the grand
potential for an interface having an arbitrarily curved geometry,

\section{Free Energy of a Curved Interface }
\label{sn:F-E-C-I}

The free energy of the interfacial region, given by Eq. (\ref{sup}),
in explicit form writes
\begin{eqnarray}
\Omega_{S} &=& {1\over 2} \int d\vec{r} \int d\vec{r}\,'\int_{0}^{1}
d\lambda \hat{n}_{\alpha}(\vec{r}) \hat{n}_{\beta}(\vec{r})
\tilde{\omega}(|\vec{r}\,'|) r'_{\alpha} \rho_0(\vec{r}-(1-\lambda)
\vec{r}\,')\nabla_{\beta} \rho_0(\vec{r}+\lambda\vec{r}\,')\nonumber \\
&-& \frac{1}{2}\int d\vec{r} \hat{n}_{\alpha}(\vec{r})
\hat{n}_{\beta}(\vec{r}) \nabla_{\nu}\int d\vec{r}\,'
\int^{1}_{0}d\lambda\rho_0(\vec{r}-(1-\lambda)\vec{r}\,')
\tilde{\omega} (|\vec{r}\,'|)\nonumber \\ &\times& r'_{\beta}
[r'_{\alpha}\nabla_{\nu}\rho_0(\vec{r}+\lambda\vec{r}\,')
-r'_{\nu}\nabla_{\alpha}\rho_0(\vec{r}+\lambda\vec{r}\,')].
\label{grapodir}
\end{eqnarray}
By performing some manipulations this expression may be compacted to
the form~\cite{segovia}
\begin{eqnarray}
\Omega_{S} &=& {1\over 2} \int d\vec{r}
\int d\vec{r}\,'\int_{0}^{1} d\lambda
\rho_0(\vec{r}-(1-\lambda)\vec{r}\,')\tilde{\omega} (|\vec{r}\,'|)
\nabla_{\alpha}\rho_0(\vec{r}+ \lambda \vec{r}\,') [r'_{\beta}
\hat{n}_{\alpha}(\vec{r})\hat{n}_{\beta}(\vec{r}) \nonumber \\
& - & 2\lambda r'_{\beta}\hat{n}_{\beta}(\vec{r})r'_{\nu}\nabla_{\alpha}
\hat{n}_{\nu}(\vec{r})+\lambda r'_{\beta}\hat{n}_{\alpha}(\vec{r})
r'_{\nu}\nabla_{\nu} \hat{n}_{\beta}(\vec{r})+\lambda r'_{\beta}
\hat{n}_{\beta}(\vec{r})r'_{\nu} \nabla_{\nu} \hat{n}_{\alpha}(\vec{r})],
\nonumber\\
\label{grapo}
\end{eqnarray}
which has been previously used to describe interfaces having planar,
spherical, and cylindrical geometry~\cite{segovia}. Direct calculation
of the grand potential in each of these geometries leads to the
equation
\begin{equation}
\Omega_{S}= -\frac{1}{4}\int d\vec{r}\int
d\vec{r}\,' \int_0^\infty ds\,\nabla\rho_0(\vec{r}\,)\cdot\nabla'
\rho_0(\vec{r}\,')\tilde{w}(s+(\vec{r}-\vec{r}\,')^2),
\label{grapoge}
\end{equation}
the difference being contained only in the density profile.\\


The first goal in this paper is to prove that Eq. (\ref{grapoge})
is satisfied for an arbitrarily curved interface, and we
concentrate on this for the remaining of this section. Instead of
starting from Eq. (\ref{grapo}), we prefer to manipulate expression
(\ref{grapodir}), introduce the linear change of variables
\begin{eqnarray}
\vec{r}^{\,(1)}  &=&
\vec{r}+\lambda \vec{r}\,',\label{ecr1}\\
\vec{r}^{\,(2)}  &=&  \vec{r}\,', \label{ecr2}
\end{eqnarray}
and use the relationship
%
\begin{equation}
{r}^{\,(2)}_{\nu}{\nabla}^{\,(1)}_{\nu}
\left[\hat{n}_{\alpha}\hat{n}_{\beta}
\left(\vec{r}^{\,(1)}-\lambda\vec{r}^{\,(2)}\right)^2\right]
= {\partial \over \partial\lambda}
\left[\hat{n}_{\alpha}\hat{n}_{\beta}
\left(\vec{r}^{\,(1)}-\lambda\vec{r}^{\,(2)}\right)^2\right],
\end{equation}
%
%
with the consideration of the symmetry with respect to the exchange of
indices $\alpha$ and $\beta$ in this equation, and that
the profile is constant at $\pm \infty$, to obtain
\begin{eqnarray}
{\Omega}_{S}& = &  -\frac{1}{2}\int d\vec{r}
\int d\vec{r}\,'  \rho_0(\vec{r})\tilde{w}(|\vec{r}\,'|)
\hat{n}_{\alpha}(\vec{r}\,)\hat{n}_{\beta}(\vec{r}\,)r_{\beta}'
{\nabla}_{\alpha}\rho_0(\vec{r}+\vec{r}\,')  \nonumber\\
&-& \frac{1}{2}\int d\vec {r}\int d\vec{r}\,' \int_0^1
d\lambda\lambda\tilde{w}(|\vec{r}\,'|)r_{\beta}'\,r_{\alpha}'
\hat{n}_{\alpha}(\vec{r}\,)\hat{n}_{\beta}(\vec{r}\,) \nonumber\\
&\times& {\nabla}_{\nu}\rho_0(\vec{r}-(1-\lambda)\vec{r}\,')
{\nabla}_{\nu}\rho_0(\vec{r}+\lambda\vec{r}\,') \nonumber\\
&-& \frac{1}{2}\int d\vec {r}\int d\vec{r}\,' \int_0^1
d\lambda\lambda\tilde{w}(|\vec{r}\,'|)r_{\beta}'\,r_{\alpha}'
\hat{n}_{\alpha}(\vec{r}\,)\hat{n}_{\beta}(\vec{r}\,) \nonumber\\
&\times& \rho_0(\vec{r}-(1-\lambda)\vec{r}\,')
{\nabla}_{\nu} {\nabla}_{\nu}\rho_0(\vec{r}+\lambda\vec{r}\,').
\label{grapmo3}
\end{eqnarray}
The last two integrals in this expression have to be manipulated to
eliminate the arbitrary parameter $\lambda$.
%
%
%
%
%
%
With this in mind we define a system of semi-orthogonal coordinates,
having unit vectors $(\hat{n}, \hat{t}_{1}, \hat{t}_{2})$, and express each
of the vectors $\vec{r}$ and $\vec{r}\,'$ on its own basis. That is,
$\vec{r}= r_{n}\hat{n}(\vec{r})+r_{t_{1}}\hat{t}_{1}+r_{t_{2}}\hat{t}_{2}$ and
$\vec{r}\,'=r_{n'}\hat{n}'(\vec{r}\,')+r_{t'_{1}}\hat{t}_{1}+r_{t'_{2}}\hat{t}_{2}$.
However, vector $\vec{r}\,'$ can also be expressed on the basis of vector
$\vec{r}$ in the following manner:
$\vec{r}\,'=(\vec{r}\,'\cdot \hat{n}(\vec{r}))\hat{n}(\vec{r})+(\vec{r}\,'
\cdot \hat{t}_{1})\hat{t}_{1}+ (\vec{r}\,'\cdot \hat{t}_{2})\hat{t}_{2}
= r'_{n}\hat{n}(\vec{r}) +r'_{t_{1}}\hat{t}_{1}+r'_{t_{2}}\hat{t}_{2}$.
We now go on to manipulating expressions so as to eliminate $\lambda$.
To achieve it we define an auxiliary function
$W(r_{n}'^2+r_{t_{1}}'^2+r_{t_{2}}'^2)$, related to
$\tilde{w}(|r_{n}'^2+r_{t_{1}}'^2+r_{t_{2}}'^2|)$ via
\begin{equation}
\frac{\partial}{\partial r_{n}'}W(r_{n}'^2+r_{t_{1}}'^2+r_{t_{2}}'^2)
=r_{n}'\tilde{w}(|r_{n}'^2+r_{t_{1}}'^2+r_{t_{2}}'^2|).
\label{truco}
\end{equation}
This relationship is employed to eliminate any power of $r'_{n}$ within the
integrals. After some more manipulations we get to the different expression
for the grand potential
\begin{eqnarray}
\Omega_{S} & = &  -\frac{1}{2}\int d\vec{r}
\int d\vec{r}\,'  \rho_0(\vec{r})\tilde{w}(|\vec{r}\,'|)
\hat{n}_{\alpha}(\vec{r}\,)\hat{n}_{\beta}(\vec{r}\,)r_{\beta}'
{\nabla}_{\alpha}\rho_0(\vec{r}+\vec{r}\,') \nonumber\\  &-&
\frac{1}{4}\int d\vec{r}\int d\vec{r}\,'\int_0^1
d\lambda\lambda \int_0^\infty
dt\,\tilde{w}(t+r_{n}'^2+r_{t_{1}}'^2+r_{t_{2}}'^2)\nonumber\\
& \times & {\nabla}_{\nu}
\rho_0(\vec{r}-(1-\lambda)\vec{r}\,'){\nabla}_{\nu}
\rho_0(\vec{r}+\lambda\vec{r}\,') \nonumber\\  &-&
\frac{1}{4}\int d\vec{r}\int d\vec{r}\,'\int_0^1
d\lambda\lambda \int_0^\infty
dt\,\tilde{w}(t+r_{n}'^2+r_{t}'^2) \nonumber\\
& \times & \rho_0(\vec{r}-(1-\lambda)\vec{r}\,')
{\nabla}_{\nu}{\nabla}_{\nu}
\rho_0(\vec{r}+\lambda\vec{r}\,') \nonumber\\  &-&
\frac{1}{8}\int d\vec{r}\int d\vec{r}\,'\int_0^1
d\lambda\lambda \int_0^\infty dt\int_0^\infty
dt'\,\tilde{w}(t+t'+r_{n}'^2+r_{t_{1}}'^2+r_{t_{2}}'^2) \nonumber\\
& \times & \frac{\partial^2}{\partial r_{n}'^2}[{\nabla}_{\nu}
\rho_0(\vec{r}-(1-\lambda)\vec{r}\,'){\nabla}_{\nu}
\rho_0(\vec{r}+\lambda\vec{r}\,')] \nonumber \\  &-&
 \frac{1}{8}\int d\vec{r}\int d\vec{r}\,'\int_0^1
d\lambda\lambda \int_0^\infty dt\int_0^\infty
dt'\,\tilde{w}(t+t'+r_{n}'^2+r_{t}'^2) \nonumber\\
& \times & \frac{\partial^2}{\partial r_{n}'^2}
[\rho_0(\vec{r}-(1-\lambda)\vec{r}\,'){\nabla}_{\nu}
{\nabla}_{\nu}\rho_0(\vec{r}+\lambda\vec{r}\,')], \label{grapomo3}
\end{eqnarray}
which may now be simplified by considering each of the terms separately.
As it is shown in Appendix \ref{sn:details}, the last four integrals
cancel one another in pairs, yielding the result for the grand potential
\begin{equation}
\Omega_{S}= -\frac{1}{2}\int d\vec {r}\int d\vec{r}\,'
\rho_0(\vec{r})\tilde{w}(|\vec{r}\,'|)\hat{n}_{\alpha}(\vec{r}\,)
\hat{n}_{\beta}(\vec{r}\,)r_{\beta}'{\nabla}_{\alpha}
\rho_0(\vec{r}+\vec{r}\,').
\label{grapmo4}
\end{equation}
The task now is to show the equivalence between this and Eq.
(\ref{grapoge}). We start by carrying out further manipulations using
(\ref{truco}) and introducing a change of variables of the form
(\ref{ecr1})--(\ref{ecr2}), but now being
%
%
%
%
%
\begin{eqnarray}
\vec{r}^{\,(1)}  &=&  \vec{r}+\vec{r}\,',\label{ecr1p}\\
\vec{r}^{\,(2)}  &=&  \vec{r}.\label{ecr2p} \label{transf1}
\end{eqnarray}
Thus, the final result obtained for the microscopic expression for the
grand potential of an arbitrarily curved interface is
%
%
\begin{eqnarray}
\Omega_{S} &=& -\frac{1}{4}\int d\vec {r}^{\,(1)}\int d\vec {r}^{\,(2)}
\int _{0}^{\infty}dt \tilde{w}(t+(\vec{r}_n^{\,(1)}-\vec{r}_n^{\,(2)})^{2}
+ (\vec{r}_{t}^{\,(1)}-\vec{r}_{t}^{\,(2)})^{2}) \nonumber\\
&\times& \partial_{n}^{(1)}\rho_0(\vec{r}^{\,(1)})
\partial_{n}^{(2)}\rho_0(\vec{r}_n^{\,(2)}),
\label{grapomo5}
\end{eqnarray}
where each vector has been written as $\vec{r}^{\,(i)}=r_n^{\,(i)}
\hat{n}^{(i)} +r_{t_1} \hat{t}_{1}^{\,(i)}+r_{t_2} \hat{t}_{2}^{\,(i)}$.\\

Notice that this expression is equal to (\ref{grapoge}), which was
to be proved. This is one of the most general results for the grand
potential that represents the free energy of the interfacial region.
The result is simple and exact. Given the interaction potential, it
only depends on the density profile.

\section{Local Approximation on the Free Energy }
\label{sn:L-A-I}

Although the previous result is simple and exact, it is not the
appropriate form to carry out comparisons with other works. In this sense,
it results convenient to get an approximation as a reference value. We
start by choosing a common basis to express all vectors. If the basis is
that of vector $\vec{r}^{\,(1)}$, vector $\vec{r}^{\,(2)}$ can be written as
$\vec{r}^{\,(2)}= (\vec{r}^{\,(2)}\cdot \hat{n}^{(1)})\hat{n}^{(1)}
+(\vec{r}^{\,(2)}\cdot \hat{t}_{1}^{\,(1)})\hat{t}_{1}^{\,(1)}
+(\vec{r}^{\,(2)}\cdot \hat{t}_{2}^{\,(1)})\hat{t}_{2}^{\,(1)}
= r_{n_2}^{(1)}\hat{n}^{(1)}+r_{t_{12}}^{(1)}\hat{t}_{2}^{\,(1)}
+r_{t_{22}}^{(1)}\hat{t}_{2}^{\,(1)}$. By introducing these into
(\ref{grapomo5}) we find
\begin{eqnarray}
\Omega_{S} &=& \frac{1}{4}\int_{0}^{\infty} d r_{n}^{(1)}\int d
{S}^{(1)}\int_{0}^{\infty} d r_{n}^{(2)}\int d S^{(2)} \int
_{0}^{\infty}dt\nonumber\\
&\times& \tilde{w}(t+(r_n^{(1)}-\vec{r}^{\,(2)}\cdot
\hat{n}^{(1)})^{2}+ (\vec{r}_{t}^{\,(1)}-\vec{r}_{t}^{\,(2)})^{2})
\nonumber\\ &\times& \hat{n}^{(1)} \cdot
\hat{n}^{(2)}\partial_{n}^{(1)}\rho_0(\vec{r}^{\,(1)})
\partial r_{n2}^{(1)}\rho_0(\vec{r}^{\,(2)}\cdot \hat{n}^{(1)}),
\label{grapomo6}
\end{eqnarray}
where $\partial r_{n2}^{(1)}={\partial}/{\partial r_{n_2}^{(1)}}$
and each element of volume has been written as
$d\vec {r}^{\,(i)}=dr_{n}^{(i)}dS^{(i)}$ with
$\vec{r}_t^{\,(i)}$ two-dimensional vectors on $S^{(1)}$ and $S^{(2)}$
respectively.\\

The expression for the grand potential that we have constructed so far
contains the description of the whole interfacial region, independently
of its details. The microscopic expression is exact for this model; exact
in the sense that the density profile contains all the information of the
geometry under consideration. In order to obtain microscopic expressions
for interfacial properties it is necessary to fix the Gibbs dividing
surface. The choice in this work is through the approximation of the
density profile as a step function. For a more general profile the
procedure becomes more complicated. We will consider that case in a
future study.\\

The density profile is thus approximated as
\begin{equation}
\rho_0(\vec{r})= \rho_0(\vec{r}_n^{\,(i)})=
\rho_{0l}\Theta(r_{n0}^{(i)}-r_n^{(i)})
+\rho_{0v}\Theta(r_{n}^{(i)}-r_{n0}^{(i)}),
\end{equation}
where $r_{n0}^{(i)}$ is the radius of the Gibbs dividing surface.
We have
\begin{equation}
\partial_n^{(i)}\rho_0( r_n^{\,(i)})=-\Delta\rho_0\,
\delta(r_n^{(i)}-r_{n0}^{(i)}),
\end{equation}
with $\Delta\rho_0 = \rho_{0v}-\rho_{0l}$. Evaluation of the
derivatives of the density profile is obvious. The result we
obtain is
\begin{equation}
\Omega_{S}= \frac{1}{4}( \Delta \rho_0 )^{2}\int d S^{(1)}\int d S^{(2)}
\hat{n}^{(1)} \cdot \hat{n}^{(2)}\int_{0}^{\infty}dt
\tilde{w}(t+(r_{n0}^{(1)}-r_{n0}^{(2)})^{2}+(\vec{r}_{t}^{\,(1)}
-\vec{r}_{t}^{\,(2)})^{2}),
\label{grapmo5}
\end{equation}
To proceed we approximate the surface representing the interface as a
paraboloid. This is possible as long as the radius of this surface is
very large compared to the range of the interaction potential. Under
this assumption, the surface can be approximated as a plane with
corrections~\cite{balian, duplantier}.\\

\begin{figure}[htp]
\begin{center}
\includegraphics[width=3.5in]{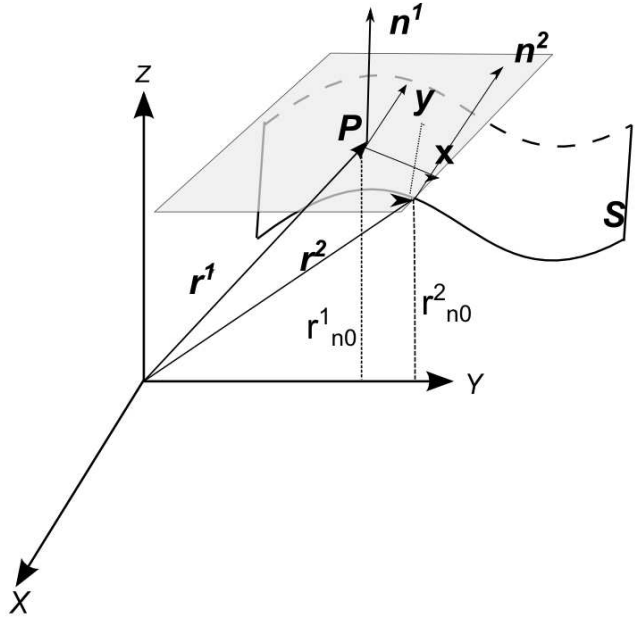}
\caption{This schematic picture shows the local approximation for
the surface $S$ about a point $P$. Points $P$ and $Q$ are localized
by vectors $\vec{r}^{\,(1)}$ and $\vec{r}^{\,(2)}$ respectively, and
the normal vectors to the surface at those points are $\hat{n}^{(1)}$
and $ \hat{n}^{(2)}$. Point $P$ is chosen as the origin of the local
coordinate system whereas $Q$ is outside the tangent plane. Its
projection onto this tangent plane has coordinates $(x,y)$.
The radius of localization of the Gibbs dividing surface is at
$r_{n0}^{(1)}$. The distance from $Q$ to the local plane (projection
of $\hat{n}^{(2)}$ onto the direction $\hat{n}^{(1)}$) is
$r_{n0}^{(1)}-r_{n0}^{(2)}=z$, with $z$ seen from the local system
as a paraboloid.}
\label{surface}
\end{center}
\end{figure}

We now build a local coordinate system in the neighborhood of a point
$P$ of the Gibbs dividing surface. This arbitrary point is localized
on the surface by vector $\vec{r}^{\,(1)}$. The normal vector at that
point is $\hat{n}^{(1)}$ and we chose the $z$ axis pointing in this
direction. Thus the coordinates $x$ and $y$ lie on the plane tangent
to the surface at $P$ and point along the directions of the principal
radii of curvature. At point $P$, $\vec{r}_{t}^{\,(1)}=(0,0)$ and
$\hat{n}^{(1)}= \hat{k}$, so that the Gibbs dividing surface is
localized at height $r_{n0}^{(1)}$. Vector $\vec{r}^{\,(2)}$ localizes
another point $Q$ on the dividing surface, close to $P$ but outside
the local plane, at a distance $z=r_{n0}^{(1)}-r_{n0}^{(2)}$ from the
plane (see Fig. \ref{surface}). If $z$ is measured from the local
system, it has the value
\begin{equation}
z={1\over 2}\left({x^{2}\over R_{1}}+ {y^{2}\over R_{2}}\right)+\cdots
\end{equation}
As vector $\vec{r}^{\,(2)}$ is not parallel to the $z$ axis of the
coordinate system, the normal vector at point $Q$ is given by
\begin{equation}
\hat{n}^{(2)}={\left(-{x\over R_{1}}, -{y\over R_{2}}, 1\right)\over
\left[1+({x\over R_{1}})^{2}+({y\over R_{2}})^{2}\right]^{1\over 2}}.
\end{equation}
The metric in this coordinate system is $g=1+ [\nabla
z(x,y)]^{2}=1+ {x^{2}\over R^{2}_{1}}+{y^{2}\over R^{2}_{2}} $,
where $\frac{x}{R_{1}}\ll 1$ and $\frac{y}{R_{2}}\ll 1$. The
surface element in the local system is $dS^{(2)}= g^{1\over 2} dx
dy$. The scalar product of the normals is $\hat{n}^{(1)} \cdot
\hat{n}^{(2)}= g^{-{1\over 2}}$. We incorporate the effect of the local
approximation into the interaction potential to obtain the free
energy of the interfacial region
%
%
%
%
\begin{eqnarray}
\Omega_{S}&=&-\frac{(\Delta\rho_0)^{2}}{4}\int dS^{(1)}
\int _{-\infty}^{\infty} dx \int _{-\infty}^{\infty}dy
\int^{\infty}_{0}dt \nonumber \\
& & {} \times \tilde{\omega}
\left(t+x^{2}+y^{2}+{1\over 4}\left({x^{2}\over R_{1}}
+{y^{2}\over R_{2}}\right)^{2} +\cdots \right).
\label{extra}
\end{eqnarray}
By expanding the interaction potential to first order about
$t+x^{2}+y^{2}$ and evaluating the integrals on the coordinates $x$
and $y$, one gets to a microscopic expression for the grand
potential in terms of the principal radii of curvature
%
%
%
%
%
\begin{eqnarray}
\Omega_{S}  & = &  - \int dS
\left \{ {(\Delta\rho_0)^{2} \pi \over 2}
\int_{0}^{\infty} dr \tilde \omega(r^{2}) r^{3} \right \}
\nonumber \\
& + & \int dS \bigg({1\over R_{1}^{2}}+ {1\over R_{2}^{2}}
+{2\over 3R_{1}R_{2}} \bigg) \left \{ {3(\Delta\rho_0)^{2}
\pi \over 64} \int_{0}^{\infty} dr \tilde \omega(r^{2}) r^{5}
\right\} +\cdots
\label{otro}
\end{eqnarray}
where we have put $dS^{(1)}= dS$. On the other hand, from the definitions
of the mean ($H$) and Gaussian ($K$) curvatures one finds that
\begin{equation}
4H^{2}-{4\over 3}K= {1\over R_{1}^{2}}+ {1\over R_{2}^{2}}
+{2\over 3R_{1}R_{2}}.
\end{equation}
By introducing this into (\ref{otro}), we get to the most general
microscopic expression for the grand potential within this approximation
level; consistent with the Helfrich prediction~\cite{Helfrich} for the
free energy
%
\begin{equation}
\Omega_{S}=-\int dS\bigg[\frac{\pi(\Delta\rho_0)^{2}}{2}
\int^{\infty}_{0}dr r^{3}\,\tilde{\omega}(r^{2})
-\frac{3 \pi(\Delta\rho_0)^{2}}{64}\left(4H^{2}-{4\over 3}K\right)
\int dr r^{5}\,\tilde{\omega}(r^{2})\bigg]. \label{real}
\end{equation}
which is also in agreement with previous results in the
planar~\cite{Triezenberg, blokhouis, segovia}, spherical~\cite{segovia},
and cylindrical~\cite{segovia} geometries, as it is shown in
Appendix \ref{sn:simp-geom}.

\section{Concluding Remarks}
\label{sn:concl}

There exist two relevant aspects of this work that are worthwhile
remarking. The first one concerns a rigorous proof for a general
expression, exact and simple, of the grand potential within a mean
field approximation. The second is a first principles derivation
of the Helfrich free energy, within this context, from a completely
original approach. This, in addition, confirms that the Helfrich
scheme is appropriate for the study of curved interfaces. Although
the local approximation of a surface as a plane that has been used
is appropriate for the description of weakly curved surfaces, we
observe that the result for the grand potential representing the
free energy of the system is sufficiently general. The microscopic
expression obtained is a function of the principal curvatures of
the surface, in complete agreement with previous predictions~%
\cite{Helfrich, robledo}. We also find complete consistency with
previous results for the simplest geometries, which were obtained
using a different analytic approach~\cite{segovia}. Finally, we
point out that within the step function approximation for the density
profile, no contribution exists from spontaneous curvature to the
free energy. We shall study the problem of an arbitrarily curved
interface for a more general density profile in a future publication.\\

\section*{Acknowledgments}
\label{sn:ack}

The authors wish to thank V.~Romero-Roch\'{\i}n
for helpful comments and stimulating discussion. This work was
supported partially by PFICA-UJAT under contract No.
UJAT-2009-C05-61 and by PROMEP-MEXICO under contract UJAT-CA-15.
J.A.S. akcnowledges financial support from PROMEP, project UAM-PTC-196.

\appendix

\section{Simplifications on the Grand Potential}
\label{sn:details}

In this appendix we summarize the main steps to simplify Eq.
(\ref{grapomo3}). We start by writing it in the form
\begin{equation}
\Omega_{S}= -\frac{1}{2}\int d\vec {r}\int d\vec{r}\,'
\rho_0(\vec{r})\tilde{w}(|\vec{r}\,'|)\hat{n}_{\alpha}(\vec{r}\,)
\hat{n}_{\beta}(\vec{r}\,)r_{\beta}'{\nabla}_{\alpha}
\rho_0(\vec{r}+\vec{r}\,')+T_{1}+T_{2}+T_{3}+T_{4},
\label{grapomo3-1}
\end{equation}
where the $T_{i}$, with $i=1,\ldots,4$, have been defined as the
integrals appearing in (\ref{grapomo3}) sequentially.\\

In order to simplify and easily identify each of the terms $T_{1}$,
$T_{2}$, $T_{3}$, and $T_{4}$, we introduce explicitly the change of
variables (\ref{ecr1})--(\ref{ecr2}). Under such a coordinate
transformation the term $T_{1}$ reads
\begin{equation}
T_{1}= -{1\over 4}\int d \vec{r}^{\,(1)} \int d \vec{r}^{\,(2)}
\int_{0}^{1} d\lambda \lambda\int _{0}^{\infty} dt\,
\tilde\omega(t+|\vec{r}^{\,(2)}|^{2})\partial_{n}^{(1)}\rho_0
(\vec{r}^{\,(1)}-\vec{r}^{\,(2)})\partial _{n}^{(1)}\rho_0 (
\vec{r}^{\,(1)}),
\end{equation}
where the notation $\partial_{n}^{(1)}= {\partial}/{\partial
r_{n}^{(1)}}$ has been introduced. Integrating by parts with
respect to $r^{(1)}_{n}$, this now becomes
 \begin{equation}
T_{1}= {1\over 4}\int d \vec{r}^{\,(1)} \int d \vec{r}^{\,(2)}
\int _{0}^{1} d\lambda\lambda\int _{0}^{\infty} dt\,
\tilde\omega(t+|\vec{r}^{\,(2)}|^{2})\rho_0(\vec{r}^{\,(1)}-\vec{r}^{\,(2)})
\partial_{n}^{(1)2}\rho_0( \vec{r}^{\,(1)}).
\end{equation}
The other term with normal derivatives of the same order is $T_{2}$.
The effect of transformation (\ref{ecr1})--(\ref{ecr2}) into this
leads to
\begin{eqnarray}
T_{2} &=& - {1\over 4}\int d \vec{r}^{\,(1)} \int d
\vec{r}^{\,(2)} \int _{0}^{1} d\lambda \lambda\int _{0}^{\infty}
dt\,\tilde \omega(t+|\vec{r}^{\,(2)}|^{2})\rho_0(\vec{r}^{\,(1)}
-\vec{r}^{\,(2)})\partial _{n}^{(1)2}\rho_0( \vec{r}^{\,(1)})\nonumber \\
&=&  -T_{1}.
\end{eqnarray}
Therefore these terms cancel one another when substituted into the
expression for the grand potential (\ref{grapomo3-1}).\\

To simplify the terms with normal derivatives of the second order,
that is $T_{3}$ and $T_{4}$, we need to take into account the following
manipulations
\begin{equation}
{\partial \over \partial r'_{n}}= {\partial r_{n}^{(1)} \over
\partial r'_{n}} {\partial \over \partial r_{n}^{(1)}} + {\partial
r_{n}^{(2)} \over \partial r'_{n}}{\partial \over \partial
r_{n}^{(2)}}= \lambda {\partial \over \partial r_{n}^{(1)} } +
{\partial \over \partial r_{n}^{(2)}}= \lambda \partial_{n}
^{(1)}+\partial_{n} ^{(2)},
\end{equation}
from where
\begin{eqnarray}
{\partial^{2}\over \partial r^{'2}_{n}}  &=& {\partial \over
\partial r'_{n}}\left(\lambda\partial_{n} ^{(1)}
+\partial_{n}^{(2)}\right)=\lambda^{2}\partial_{n}^{(1)2}+
2\lambda
\partial_{n}^{(1)}\partial_{n}^{(2)}+\partial_{n}^{(2)2}.
\end{eqnarray}
By introducing this into $T_{3}$, we obtain
\begin{eqnarray}
T_{3} &=&  - {1\over 8}\int d \vec{r}^{\,(1)} \int d
\vec{r}^{\,(2)} \int _{0}^{1} d\lambda \lambda\int
_{0}^{\infty}dt\int _{0}^{\infty}
dt'\tilde \omega(t+t'+|\vec{r}^{\,(2)}|^{2})\nonumber\\
&\times & \left[\lambda^{2}\partial_{n}^{(1)2}+2\lambda
\partial_{n}^{(1)}\partial_{n}^{(2)}+\partial_{n}^{(2)2}\right]
\left[\partial _{n}^{(1)}\rho_0(\vec{r}^{\,(1)}-\vec{r}^{\,(2)})
\partial _{n}^{(1)}\rho_0( \vec{r}^{\,(1)})\right].
\end{eqnarray}
We need to calculate derivatives of the product within the
integrand. The first one is
\begin{eqnarray}
\partial_{n}^{(1)}[\partial_{n}^{(1)}\rho_0(\vec{r}^{\,(1)}-\vec{r}^{\,(2)})
\partial_{n}^{(1)}\rho_0(\vec{r}^{\,(1)})]   & = &
\partial_{n}^{(1)2}\rho_0(\vec{r}^{\,(1)}-\vec{r}^{\,(2)})
\partial_{n}^{(1)}\rho_0(\vec{r}^{\,(1)})\nonumber \\
& + & \partial_{n}^{(1)}\rho_0(\vec{r}^{\,(1)}-\vec{r}^{\,(2)})
\partial_{n}^{(1)2}\rho_0( \vec{r}^{\,(1)}),
\end{eqnarray}
and the second
\begin{eqnarray}
\partial_{n}^{(1)2}[\partial_{n}^{(1)}\rho_0(\vec{r}^{\,(1)}-\vec{r}^{\,(2)})
\partial_{n}^{(1)}\rho_0(\vec{r}^{\,(1)})] &=&
\partial_{n}^{(1)}[\partial_{n}^{(1)2}\rho_0(\vec{r}^{\,(1)}
-\vec{r}^{\,(2)})\partial_{n}^{(1)}\rho_0( \vec{r}^{\,(1)}) \nonumber\\
&+&\partial_{n}^{(1)}\rho_0(\vec{r}^{\,(1)}-\vec{r}^{\,(2)})\partial_{n}^{(1)2}
\rho_0( \vec{r}^{\,(1)})]\nonumber\\  &=&  \partial_{n}^{(1)3}
\rho_0(\vec{r}^{\,(1)}-\vec{r}^{\,(2)})\partial_{n}^{(1)}
\rho_0(\vec{r}^{\,(1)}) \nonumber\\ &+& 2\partial_{n}^{(1)2}
\rho_0(\vec{r}^{\,(1)}-\vec{r}^{\,(2)})\partial_{n}^{(1)2}
\rho_0(\vec{r}^{\,(1)}) \nonumber \\
&+& \partial_{n}^{(1)}\rho_0(\vec{r}^{\,(1)}-\vec{r}^{\,(2)})
\partial_{n}^{(1)3}\rho_0(\vec{r}^{\,(1)}).
\end{eqnarray}
The corresponding derivatives for $\partial_{n}^{(2)}$ are also
calculated. By introducing all these into $T_{3}$ one finds
\begin{eqnarray}
T_{3} &=& - {1\over 8}\int d \vec{r}^{\,(1)} \int d
\vec{r}^{\,(2)} \int _{0}^{1} d\lambda \lambda\int _{0}^{\infty}
dt\int _{0}^{\infty}dt'\tilde\omega(t+t'+|\vec{r}^{\,(2)}|^{2})
\nonumber\\ &\times&
\{\lambda^{2}[\partial_{n}^{(1)3}\rho_0(\vec{r}^{\,(1)}-\vec{r}^{\,(2)})
\partial_{n}^{(1)}\rho_0(\vec{r}^{\,(1)})+
\partial_{n}^{(1)2}\rho_0(\vec{r}^{\,(1)}-\vec{r}^{\,(2)})
\partial_{n}^{(1)2}\rho_0(\vec{r}^{\,(1)})\nonumber\\
&+& \partial_{n}^{(1)}\rho_0(\vec{r}^{\,(1)}
-\vec{r}^{\,(2)})\partial_{n}^{(1)3}\rho_0(\vec{r}^{\,(1)})]
+2\lambda[\partial_{n}^{(1)2}\rho_0(\vec{r}^{\,(1)}
-\vec{r}^{\,(2)})\partial_{n}^{(1)}\rho_0( \vec{r}^{\,(1)})
\nonumber \\ &+&  \partial_{n}^{(1)}
\rho_0(\vec{r}^{\,(1)}-\vec{r}^{\,(2)})\partial_{n}^{(1)2} \rho_0(
\vec{r}^{\,(1)})]+\partial_{n}^{(2)2}[\partial_{n}^{(1)}
\rho_0(\vec{r}^{\,(1)}-\vec{r}^{\,(2)})\partial _{n}^{(1)}
\rho_0(\vec{r}^{\,(1)})] \}.\nonumber\\
\end{eqnarray}
or alternatively, after an integration by parts respect to
$r_{n}^{(1)}$ in each term,
\begin{eqnarray}
T_{3} &=&  {1\over 8}\int d \vec{r}^{\,(1)} \int d \vec{r}^{\,(2)}
\int _{0}^{1} d\lambda \lambda\int _{0}^{\infty} dt\int
_{0}^{\infty}dt'\tilde\omega(t+t'+|\vec{r}^{\,(2)}|^{2})
\nonumber\\ &\times&
\{\lambda^{2}[\partial_{n}^{(1)2}\rho_0(\vec{r}^{\,(1)}-\vec{r}^{\,(2)})
\partial_{n}^{(1)2}\rho_0(\vec{r}^{\,(1)})
2\partial_{n}^{(1)}\rho_0(\vec{r}^{\,(1)}-\vec{r}^{\,(2)})
\partial_{n}^{(1)3}\rho_0(\vec{r}^{\,(1)})\nonumber \\  &+&
\rho_0(\vec{r}^{\,(1)}-\vec{r}^{\,(2)})
\partial_{n}^{(1)4}\rho_0(\vec{r}^{\,(1)}]+2\lambda\partial_{n}^{(2)}
[\partial_{n}^{(1)}\rho_0(\vec{r}^{\,(1)}-\vec{r}^{\,(2)})\partial_{n}^{(1)2}
\rho_0( \vec{r}^{\,(1)})\nonumber \\  &+& \rho_0(\vec{r}^{\,(1)}
-\vec{r}^{\,(2)})\partial_{n}^{(1)3}\rho_0( \vec{r}^{\,(1)})]
+\partial_{n}^{(2)2}[\rho_0(\vec{r}^{\,(1)}-\vec{r}^{\,(2)})
\partial_{n}^{(1)2}\rho_0( \vec{r}^{\,(1)})] \}.
\end{eqnarray}
Now we investigate the effect of the same transformation in the
expression for $T_{4}$. Direct substitution implies
\begin{eqnarray}
T_{4}  &=&  - {1\over 8}\int d \vec{r}^{\,(1)} \int d
\vec{r}^{\,(2)} \int _{0}^{1} d\lambda \lambda\int _{0}^{\infty}
dt\int _{0}^{\infty}dt'\tilde \omega(t+t'+|\vec{r}^{\,(2)}|^{2})
\nonumber\\  &\times &
\left[\lambda^{2}\partial_{n}^{(1)2}+2\lambda
\partial_{n}^{(1)}\partial_{n}^{(2)}+\partial_{n}^{(2)2}\right]
\left[\rho_0(\vec{r}^{\,(1)}-\vec{r}^{\,(2)})\partial _{n}^{(1)2}
\rho_0( \vec{r}^{\,(1)})\right].
\end{eqnarray}
Although this compact form appears simple, we need to carry out
further simplifications to identify similar terms. We start by
calculating the first derivative
\begin{eqnarray}
\partial_{n}^{(1)}[\rho_0(\vec{r}^{\,(1)}-\vec{r}^{\,(2)})
\partial^{(1)2}\rho_0(\vec{r}^{\,(1)})] &=& \partial_{n}^{(1)}
\rho_0(\vec{r}^{\,(1)}-\vec{r}^{\,(2)})\partial_{n}^{(1)2}
\rho_0( \vec{r}^{\,(1)}) \nonumber\\
&+& \rho_0(\vec{r}^{\,(1)}-\vec{r}^{\,(2)})
\partial_{n}^{(1)3}\rho_0( \vec{r}^{\,(1)}).
\end{eqnarray}
and then a further derivative of this quantity in the same
direction $\partial_{n}^{(1)}$, which yields
\begin{eqnarray}
\partial_{n}^{(1)2}[\rho_0(\vec{r}^{\,(1)}-\vec{r}^{\,(2)})
\partial_{n}^{(1)2}\rho_0(\vec{r}^{\,(1)})] &=&
\partial_{n}^{(1)}[ \partial_{n}^{(1)}\rho_0(\vec{r}^{\,(1)}
-\vec{r}^{\,(2)})\partial_{n}^{(1)2}\rho_0( \vec{r}^{\,(1)})
\nonumber\\
&+& \rho_0(\vec{r}^{\,(1)}-\vec{r}^{\,(2)})\partial_{n}^{(1)3}
\rho_0(\vec{r}^{\,(1)})]\nonumber
\\  &=&
\partial_{n}^{(1)2}\rho_0(\vec{r}^{\,(1)}-\vec{r}^{\,(2)})
\partial_{n}^{(1)2}\rho_0(\vec{r}^{\,(1)}) \nonumber\\
&+& 2\partial_{n}^{(1)}
\rho_0(\vec{r}^{\,(1)}-\vec{r}^{\,(2)})\partial_{n}^{(1)3}
\rho_0(\vec{r}^{\,(1)})\nonumber \\
 &+& \rho_0(\vec{r}^{\,(1)}-\vec{r}^{\,(2)})
\partial_{n}^{(1)4}\rho_0(\vec{r}^{\,(1)}).
\end{eqnarray}
By substituting we find
\begin{eqnarray}
T_{4} &=& - {1\over 8}\int d \vec{r}^{\,(1)} \int d
\vec{r}^{\,(2)} \int _{0}^{1} d\lambda \lambda\int _{0}^{\infty}
dt\int _{0}^{\infty}dt'\tilde\omega(t+t'+|\vec{r}^{\,(2)}|^{2})
\nonumber\\
&\times&
\{\lambda^{2}[\partial_{n}^{(1)2}\rho_0(\vec{r}^{\,(1)}-\vec{r}^{\,(2)})
\partial_{n}^{(1)2}\rho_0(\vec{r}^{\,(1)})
+2\partial_{n}^{(1)}\rho_0(\vec{r}^{\,(1)}-\vec{r}^{\,(2)})
\partial_{n}^{(1)3}\rho_0(\vec{r}^{\,(1)})
\nonumber \\  &+& \rho_0(\vec{r}^{\,(1)}-\vec{r}^{\,(2)})
\partial_{n}^{(1)4}\rho_0(\vec{r}^{\,(1)})]
+2\lambda\partial_{n}^{(2)}[\partial_{n}^{(1)}\rho_0(\vec{r}^{\,(1)}
-\vec{r}^{\,(2)})\partial_{n}^{(1)2}\rho_0( \vec{r}^{\,(1)})
\nonumber \\  &+& \rho_0(\vec{r}^{\,(1)}
-\vec{r}^{\,(2)})\partial_{n}^{(1)3}\rho_0( \vec{r}^{\,(1)})]
+\partial_{n}^{(2)2}[\rho_0(\vec{r}^{\,(1)}
-\vec{r}^{\,(2)})\partial _{n}^{(1)2}\rho_0( \vec{r}^{\,(1)})]\}
\nonumber\\ &=& -T_{3}.
\end{eqnarray}
That is, also the terms $T_{3}$ and $T_{4}$ cancel one another.

\section{Simplest Geometries}
\label{sn:simp-geom}

To make contact with previous results we consider here the simplest
geometries. For the planar surface $R_{1}= R_{2}= \infty$ so that
$4H^{2}-{4\over 3}K = 0$ and thus Eq. (\ref{real}) simplifies to
\begin{equation}
\Omega_{S}=-\int dS\bigg[ \frac{\pi(\Delta\rho_0)^{2}}{2}
\int^{\infty}_{0}dr r^{3}\,\tilde{\omega}(r^{2}) \bigg].
\label{realpla}
\end{equation}
That is, the only contribution to the free energy comes from
the surface tension term, in agreement with previous
results~\cite{Triezenberg, blokhouis, segovia}.\\

For a spherical interface $R_{1}= R_{2}= R$, $H={1\over R}$, and
$K= {1\over R^2}$, so that $4H^{2}-{4\over 3}K= {8\over 3R^{2}}$.
From (\ref{real}) one gets to the corresponding free energy
\begin{equation}
\Omega_{S}=-\int dS\bigg[\frac{\pi(\Delta\rho_0)^{2}}{2}
\int^{\infty}_{0}dr r^{3}\,\tilde{\omega}(r^{2}) -\frac{
\pi(\Delta\rho_0)^{2}}{8R^{2}} \int dr
r^{5}\,\tilde{\omega}(r^{2})\bigg]. \label{realesf}
\end{equation}
Finally, for the cylindrical interface $R_{1}= \infty$ and
$R_{2}=R$, so that $4H^{2}-{4\over 3}K= {1\over R^2}$.
In this case the free energy is
\begin{equation}
\Omega_{S}=-\int dS\bigg[\frac{\pi(\Delta\rho_0)^{2}}{2}
\int^{\infty}_{0}dr r^{3}\,\tilde{\omega}(r^{2}) -\frac{3
\pi(\Delta\rho_0)^{2}}{64R^{2}} \int dr
r^{5}\,\tilde{\omega}(r^{2})\bigg]. \label{realcil}
\end{equation}
These last two expressions are consistent with previous
results~\cite{segovia}.


\begin{thebibliography}{00}


\bibitem{evans} R.~Evans, Adv. Phys. {\bf 28}, 143 (1979).

\bibitem{widom} J.~S.~Rowlinson and B.~Widom,
{\it Molecular Theory of Capillarity} (Clarendon, Oxford, 1982).

\bibitem{tolman} R.~C.~Tolman, J. Chem. Phys. {\bf 17}, 333 (1949).

\bibitem{dietrich} S.~Dietrich and M.~Napiorkowski,
Physica A {\bf 177}, 437 (1991);\\ M.~Napiorkowski and S.~Dietrich,
Phys. Rev. E {\bf 47}, 1836 (1993);\\ M.~Napiorkowski and S.~Dietrich,
Z. Phys. B {\bf 97}, 511 (1995);\\ K.~R.~Mecke and S.~Dietrich,
Phys. Rev. E {\bf 59}, 6766 (1999);\\ K.~R.~Mecke and S.~Dietrich,
J. Chem. Phys. {\bf 123}, 204723 (2005).

\bibitem{Triezenberg} D.~G.~Triezenberg and R.~Zwanzig,
Phys. Rev. Lett. {\bf 28}, 1183 (1972). This result is known to have
been obtained by Yvon but he did not published it.

\bibitem{gibbs} A.~J.~M.~Yang, P.~D.~Fleming, and J.~H.~Gibbs,
J. Chem. Phys. {\bf 64}, 3732 (1976);\\ A.~J.~M.~Yang, P.~D.~Fleming,
and J.~H.~Gibbs, J. Chem. Phys. {\bf 67}, 74 (1977).

\bibitem{blokhouis} E.~M.~Blokhuis and D.~Bedeaux,
Physica A {\bf 184}, 42 (1992);\\ E.~M.~Blokhuis and D.~Bedeaux,
Mol. Phys. {\bf 80}, 705 (1993);\\ A.~E.~van Giessen, E.~M.~Blokhuis,
and D.~J.~Bukman, J. Chem. Phys. {\bf 108}, 1148 (1998).

\bibitem{Percus} J.~K.~Percus, in {\it The Liquid State of Matter:
Fluids, Simple and Complex}, E.~W.~Montroll and J.~L.~Lebowitz, eds.
(North-Holland, Amsterdam, 1982).

\bibitem{romero} V.~Romero-Roch\'{\i}n, C.~Varea, and A.~Robledo
Phys. Rev. A {\bf 44}, 8417, (1991);\\ V.~Romero-Roch\'{\i}n,
C.~Varea, and A.~Robledo, Physica A {\bf 184}, 367 (1992);\\
V.~Romero-Roch\'{\i}n, C.~Varea, and A.~Robledo, Mol. Phys.
{\bf 80}, 821 (1993);\\  V.~Romero-Roch\'{\i}n, C.~Varea, and
A.~Robledo, Phys. Rev. E {\bf 48}, 1600 (1993).

\bibitem{percus} J.~K.~Percus, J. Math. Phys. {\bf 37}, 1259 (1996).

\bibitem{romeropercus} V.~Romero-Roch\'{\i}n and J.~K.~Percus,
Phys. Rev. E {\bf 53}, 5130 (1996).

\bibitem{segovia} J\'ose~G.~Segovia-L\'opez and V\'{\i}ctor
Romero-Roch\'{\i}n, Phys. Rev. E {\bf 73}, 021601, (2006).

\bibitem{Helfrich} W.~Helfrich, Z. Naturfosch. Teil A {\bf 28},
693 (1973).

\bibitem{robledo1} C.~Varea and A.~Robledo, Physica A {\bf 220},
33 (1995);\\ C.~Varea and A.~Robledo, Mol. Phys. {\bf 85}, 477
(1995).

\bibitem{robledo} A.~Robledo and C.~Varea, Physica A {\bf 231},
178 (1996).

\bibitem{Landau-elas} L.~D.~Landau and E.~M.~Lifshitz, {\it Theory
of Elasticity} (Pergamon Press, Oxford, 1984).

\bibitem{balian} R.~Balian and C.~Bloch, Ann. Phys. (N.~Y.)
{\bf 60}, 401 (1970).

\bibitem{duplantier} Bertand~Duplantier, Raymond~E.~Goldstein,
Victor~Romero-Roch\'{\i}n, and Adriana~I.~Pesci,
Phys. Rev. Lett. {\bf 65}, 508 (1990).



\end{thebibliography}
\end{document}